\documentclass[12pt]{article}

\def\be{\begin{eqnarray}}
\def\ee{\end{eqnarray}}
\def\bq{\begin{equation}}
\def\eq{\end{equation}}
\def\ben{\begin{enumerate}}\def\een{\end{enumerate}}

\def\roughly#1{\mathrel{\raise.3ex\hbox{$#1$\kern-.75em%
\lower1ex\hbox{$\sim$}}}}

\usepackage{epsfig}

\begin{document}
\begin{titlepage}

\hfill FTUV-06-0922

 \vspace{1.5cm}
\begin{center}
\ \\
{\bf\LARGE Glueball enhancement by color de-confinement}
\\
\vspace{0.7cm} {\bf\large Vicente Vento} \vskip 0.7cm

{\it  Departamento de F\'{\i}sica Te\'orica and Instituto de
F\'{\i}sica Corpuscular}

{\it Universidad de Valencia - Consejo Superior de Investigaciones
Cient\'{\i}ficas}

{\it 46100 Burjassot (Val\`encia), Spain, }

{\small Email: Vicente.Vento@uv.es}

\end{center}
\vskip 1cm \centerline{\bf Abstract} \vskip 0.3cm

High energy heavy ion collisions lead to the formation of a strong
coupling de-confined phase in which the lightest glueballs are
numerous and stable. We analyze how their properties manifest
themselves in experimental spectra and show that they provide a
good signature for color de-confinement.
 \vspace{2cm}

\noindent Pacs: 14.80.-j, 24.80.+y, 25.75.Nq


\end{titlepage}

\section{Introduction}

\indent\indent Quantum Chromodynamics (QCD)  is the theory of the
strong interactions \cite{FritzschGellMannLeutwyler}. At low
temperatures its elementary constituents are mesons, baryons and
glueballs \cite{MinkowskiFritzsch}, this is the hadronic phase
where all states are color singlets. At very high temperatures one
expects a phase transition, called de-confinement, to take place.
The new phase was thought to be a plasma of quarks and gluons and
was named the Quark Gluon Plasma (QGP)\cite{QGP}. However, a
formulation of the dynamics in the region above the transition
temperature $T_C$, based on a description of recent experiments in
ultra relativistic heavy ion collisions \cite{RHIC}, states that,
despite de-confinement, the color Coulomb interaction between the
constituents is strong and a large number of binary (even color)
bound states, with a specific mass pattern, are formed
\cite{Shuryak}. This phase I call Strong Coulomb Phase (SCP). The
QGP phase occurs at a much higher temperature $T_{QGP} > (2-3)
T_C$.

My aim is to study the behavior of QCD in the transition from the
SCP to the hadronic phase centering my attention in the behavior
of the scalar glueballs. I am not the first to propose the
glueballs to characterize de-confinement \cite{Shuryak0} but I
single out in here a different dynamical picture . Glueballs are
bound states of gluons, the gauge bosons of QCD. This unique
structure led to an intense experimental search, since they were
first theoretically contemplated \cite{MinkowskiFritzsch}, which
has not produced a clear picture of their spectrum. Moreover, they
are expected to be broad because they mix strongly with quark
states. The lightest glueball is a $0^{++}$, which I shall label
by $g$.

Recently, I have proposed an interpretation of the scalar
particles that contemplates a rich low lying glueball spectrum
\cite{Vento}. The analysis has been modelled by $\frac{1}{N_C}$
physics on which I have also based my estimates. I was led to a
dynamical scenario where the OZI rule is broken and a low mass
glueball, $g$, arises from the mixing of a pure OZI conserving
glueball and a $\sigma$-meson, which are almost degenerate in
mass. In this scenario $g$ is narrow although  it remains hidden
in the tale of the $\sigma$-meson. Experimentally they appear as a
unique and broad resonance, the $f_0 (600)$ \cite{Vento}.

My purpose is to show that these two states, $g$ and $\sigma$,
behave in a very characteristic way across the de-confinement
transition and therefore lead to observable effects associated
with the SCP phase transition.

\section{Behavior of the $g$ with temperature}

The realization of scale symmetry in Gluodynamcis (GD), the theory
with gluons and no quarks, provides a relation between the
parameters of the lightest scalar glueball, hereafter called $g$,
and the gluon condensate \cite{Schechter,Migdal,EllisLanik},

\bq m_g^2 f_g^2 = - 4 \; <0| \; \frac{\beta(\alpha_s)}{4
\alpha_s}\; G^2\;|0>,\label{mgfg} \eq
where $f_g = <0|g|0>$ , $m_g$ the $g$ mass, and the right hand
side arises from the scale anomaly. GD provides a description for
glueballs which almost coincides with that of QCD in the limit
when the OZI rule is exactly obeyed, i.e., when decays into quarks
which require gluons are strictly forbidden \cite{Vento}.

Lattice results \cite{Lattice,Langfeld} and model calculations
\cite{Heinz,Drago,Wambach} support the traditional scenario
 \cite{Leutwyler,Miller}, that the condensate is basically
constant up to the phase transition temperature $T_C$ ($150 MeV <
T_C < 300 MeV$) and decreases slowly thereafter until it dilutes
(or evaporates) into gluons at $(2-3)  T_C $. In this regime the
mass of $g$ changes slowly across the phase transition
\cite{Heinz, Drago,Wambach} and might even increase beyond $T_C$
as the gluon binding energy decreases \cite{Shuryak} (see Fig. 1).
These results and Eq.(\ref{mgfg}) determine that $f_g$ will be
small only close to the dilution temperature when, in GD, scale
invariance is restored. However, around $T_C$, $f_g$ is sizeable
and therefore we are able to use in the scalar sector the OZI
approximation of QCD, where glueballs and mesons are almost
decoupled, and therefore the scalar glueballs of QCD behave
similarly to those of GD \cite{Vento}.

\vskip 0.5cm

\begin{figure}[htb]
\begin{minipage}[t]{7.5cm}
\hskip-0.5cm \epsfig{file=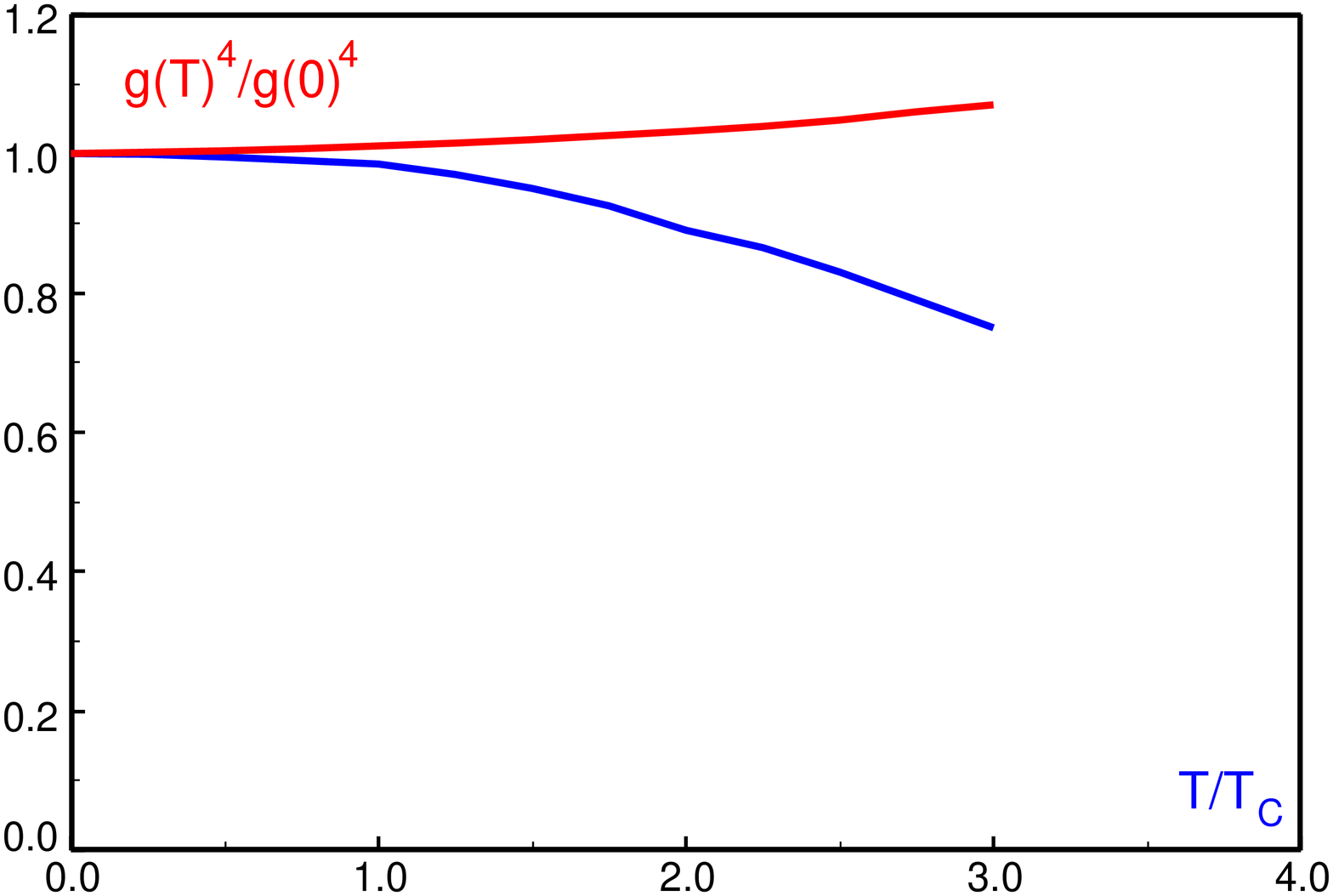,width=7.5cm,angle=0}
\end{minipage} \hskip -1.0cm\begin{minipage}[t]{7.5cm}
\epsfig{file=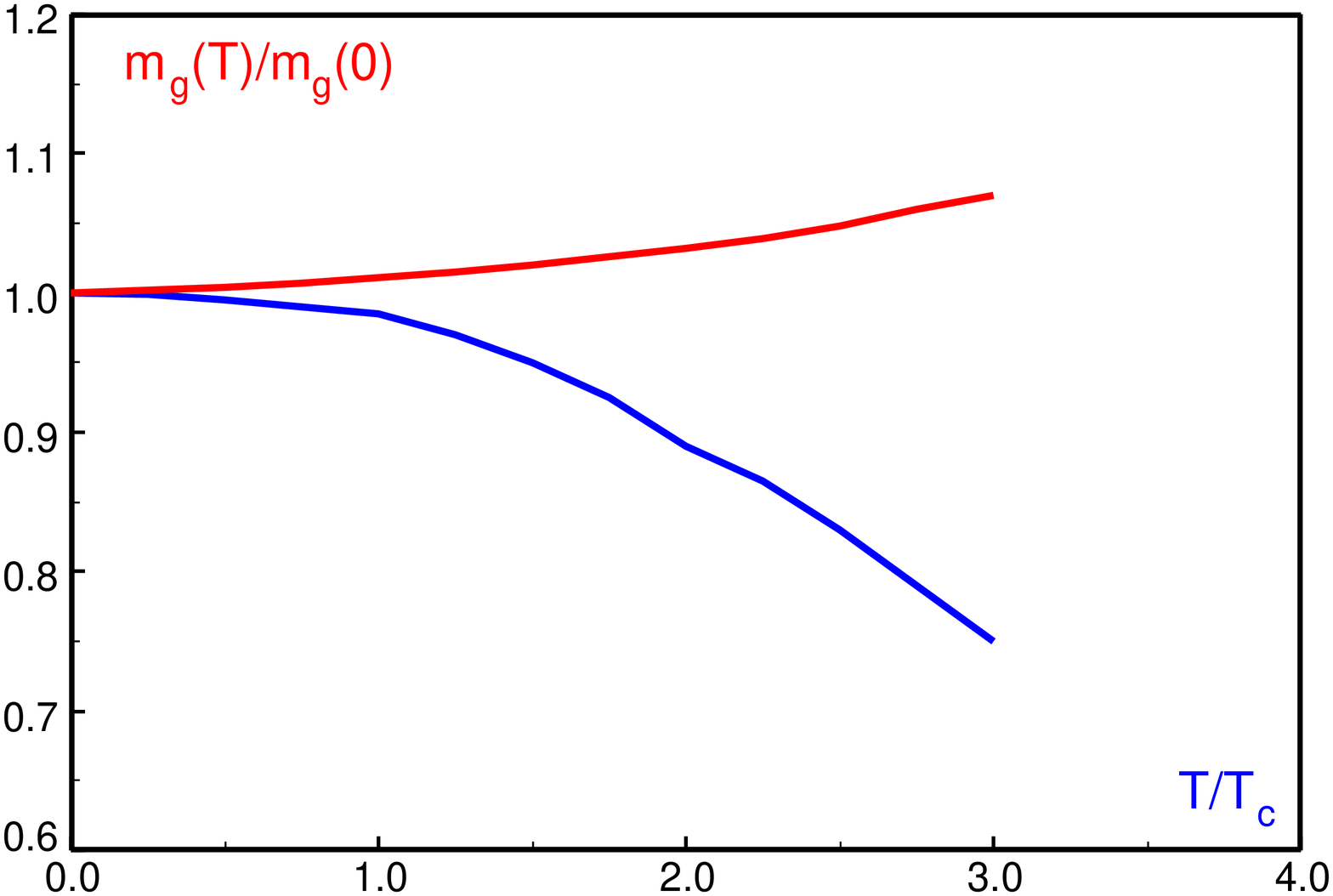,width=7.5cm,angle=0}
\end{minipage}
\caption{\small{Behavior of the gluon condensate, $g^4=
m_g^2f_g^2$, and the mass,$m_g$, of $g$ across the de-confinement
phase transition for various calculations mentioned in the text.
The allowed values for different model calculations occur between
the two lines. \label{condmass}}} \vskip 0.5cm
\end{figure}

I assume for the discussion a recent formulation of the dynamics
in the region above $T_C$, which states that despite
de-confinement the color Coulomb interaction between the
constituents is strong and a large number of binary (even color)
bound states, with a specific mass pattern, are formed
\cite{Shuryak}. With this input, the scenario I envisage for GD
goes as follows\cite{Vento1}. The strong Coulomb phase is crowded
with gluon bound states and $g$ is the lightest. As one moves
towards the dilution limit, the binding energy of these states
decreases, the gluon mass increases, and therefore the color and
singlet bound states increase their mass softly until the gluons
are liberated forming a liquid \cite{Shuryak,Polyakov,Linde}.
However, as one cools towards the confining phase, color and
singlet states decay into the conventional low lying glueballs, in
particular $g$. Thus the number of $g$'s becomes large.

Moreover, going to QCD, one realizes that above the phase
transition the multiplicity of glueball channels is larger than
below. The ratio of glueball to meson channels goes from 1 to 8
below the phase transition to 1 to 2 above \cite{Shuryak}.

Thus our first result is that the number of scalar glueballs is
much larger in SCP than in the cold world.

\vskip 0.5cm
\begin{figure}[htb]
\centerline{\epsfig{file=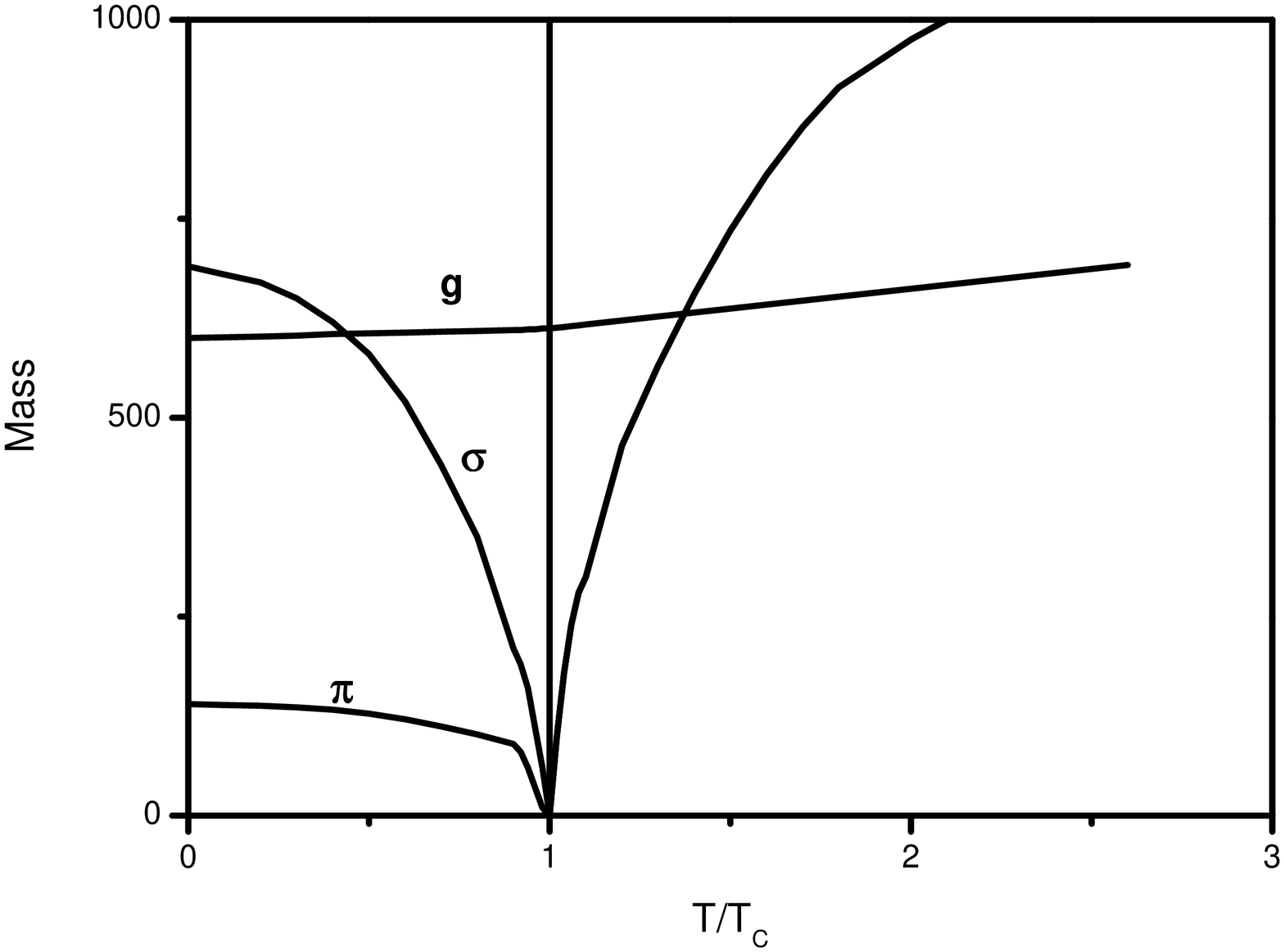,width=7.0cm,angle=270}}
\caption{\small{Behavior of the masses of $\sigma$,$\pi$ and $g$
across the QGP phase transition according to model calculations.
\label{SCP}}} \vskip 0.5cm
\end{figure}

The physical $g$ and $\sigma$ arise from the degenerate ones via
mixing,

\be g & =  & g_0 \cos{(\theta/2)} - \sigma_0 \sin{(\theta/2)} ,
\\ \sigma & =  & g_0 \sin{(\theta/2)}  + \sigma_0 \cos{(\theta/2)},
\ee
where, the physical fields are expressed in terms of the
degenerate OZI fields, $g_0$ and $\sigma_0$, and the mixing angle
$\theta$ .

The $\sigma$ meson decays into two pions or two photons
\cite{Vento},

\be \Gamma_{\sigma\rightarrow 2\pi} & \sim &  1.5 \cos ^2
{(\theta/2)} \left(\frac{m_\sigma
(\mbox{GeV})}{1\mbox{GeV}}\right)^3 \mbox{GeV} ,
\\
\Gamma_{\sigma\rightarrow 2\gamma} & \sim &  10.5 \cos ^2
{(\theta/2)} \left(\frac{m_\sigma
(\mbox{GeV})}{1\mbox{GeV}}\right)^3 \mbox{eV} .\ee
Therefore the physical scalar glueball $g$ decays, due to its
$\sigma$ component,  decays also into two pions or two photons
\cite{Vento},

\be  \Gamma_{g\rightarrow 2\pi} & \sim & 1.5 \sin ^2 {(\theta/2)}
\left(\frac{m_g (\mbox{GeV})}{1\mbox{GeV}}\right)^3 \mbox{GeV},
\\ \Gamma_{g \rightarrow 2\gamma} & \sim & 10.5
\sin ^2 {(\theta/2)} \left(\frac{m_g
(\mbox{GeV})}{1\mbox{GeV}}\right)^3 \mbox{eV}, \ee
If we assume that the $\sigma$ is the O(4) partner of the $\pi$ in
the chiral symmetry realization of QCD, its mass decreases when
approaching the phase transition, becoming degenerate with the
pion  at $T_C$ (see Fig.2). Beyond $T_C$, in the SCP, chiral
symmetry is restored, and $\pi$ and $\sigma$ remain degenerate for
$T > T_C$. Thus in the SCP the $\sigma$ can only decay in
$2\gamma$ for obvious kinematical reasons. The glueball $g$ does
not vary its mass in this region appreciably. Thus even before we
reach $T_C$, the mixing between $g$ and $\sigma$ disappears (see
Fig.2) and $g$ becomes stable around $T_C$. However, in the SCP
the mass of the $\sigma$ increases and in a certain region of $T$
it again becomes degenerate with $g$ and mixing is restored. Thus
the physical  $g$ is able to decay, once the $\sigma$ component is
attained, to  $2\gamma$.

Summarizing, in the SCP the ratio of the scalar glueballs to
mesons increases and both the $g$ and the $\sigma$ can only decay
to $2\gamma$.

\section{The cooling of the fireball}

When two heavy ions collide at ultra-relativistic energies, if the
collision is quite central, a hot region of space time is produced
called the fireball \cite{RHIC}. Let me incorporate in the cooling
of the fireball the dynamics of QCD as described above
\cite{Vento1}. My starting point is SCP with a temperature $T_C <
T < 3T_C$. This plasma is almost a perfect fluid of hadronic
matter with low viscosity and full of binary states
\cite{Shuryak}. The lowest mass $q \bar{q}$ states are the
pseudoscalar pion $\pi$ and the scalar meson $\sigma$, which are
here bound states of the strong color interaction. The lightest
glueball state is $g$. The behavior of $g$ runs together with all
other hadronic processes leading to a collective flow but, in the
OZI approximation, it can be singled out. Let me consider how the
cooling of this plasma affects the population of glueballs and
their general flow.

\vskip 0.5cm
\begin{figure}[htb]
\centerline{\epsfig{file=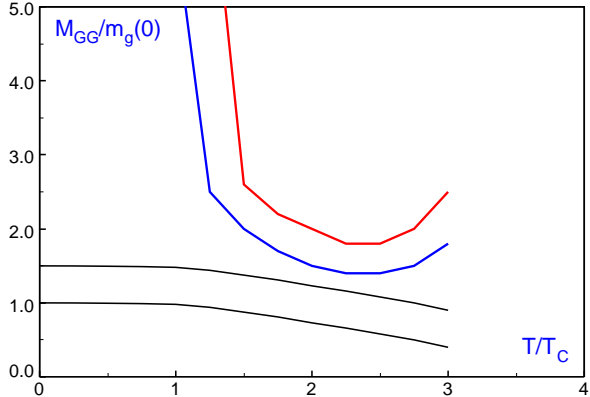,width=9.0cm,angle=0}}
\caption{\small{Behavior of the glueball masses as one approaches
the de-confinement phase transition. The mass of the color singlet
states remains basically constant while the mass of the color
states increases dramatically. \label{ggmass}}} \vskip 0.5cm
\end{figure}

As the fireball cools a ``large number" of gluonic bound states
decay by gluon emission into $g$'s. The emitted gluons form new
bound states of lower mass due to the strong color Coulomb
interaction. As we approach the confinement region the mass of the
color bound states increases and it pays off to make multiparticle
color singlet states, which decay by rearrangement into ordinary
color singlet states (see Fig.3). Since the coupling is strong and
the phase space is large, these processes take place rapidly. Thus
in no time, close to the phase transition temperature $T_C$, a
large number of scalar glueballs populate the hadronic liquid. In
our idealized OZI world they interact among themselves and with
quark matter only by multi gluon exchanges, i.e., weak long range
color Van der Waals forces. These forces allow the glueballs to be
dragged by the hadronic liquid with the flow determined by the
kinetics of the binary states from which they all proceed
\cite{Shuryak}.

\section{Experimental signatures}

I have presented all the ingredients necessary to discuss the
observational signatures. I foresee two types of signatures:
$2\gamma$ decays  and  $2\pi$ decays.

It is clear that the $\sigma$ and the $g$ will decay in the SCP
only to $2 \gamma$. The width of the $\sigma$ will be strongly
temperature dependent, due to the temperature dependence of its
mass. On the contrary the $g$ will only decay in the temperature
range where mixing takes place, i.e. when their masses are close
(see Fig.2). Besides their mass variation both widths will be
broaden by the energy of the heat bath. Thus we expect a broad
$\sigma$ meson, whose width is temperature dependent, and a narrow
$g$ whose width is temperature independent but has a temperature
threshold. The enhancement in the $g$ with respect to the hadronic
phase arises because of the larger population in the SCP as
described above, since the ratio of glueball channels to meson
channels changes from 1:8 to 1:2, and because these particles are
stable in the medium against the dominating hadronic decays. The
PHOS detector at LHC, whose capability of detecting $\pi_0$ is
impressive \cite{Kharlov} should be able to detect both the
$\sigma$ and the $g$ (see Fig.4) \footnote{I expect that the
broadening of the $\pi_0$ width is not only due to the heat bath
but also to the temperature dependence of the pion mass}.

\vskip 0.5cm
\begin{figure}[htb]
\begin{minipage}[t]{6.0cm}
\hskip-0.0cm \epsfig{file=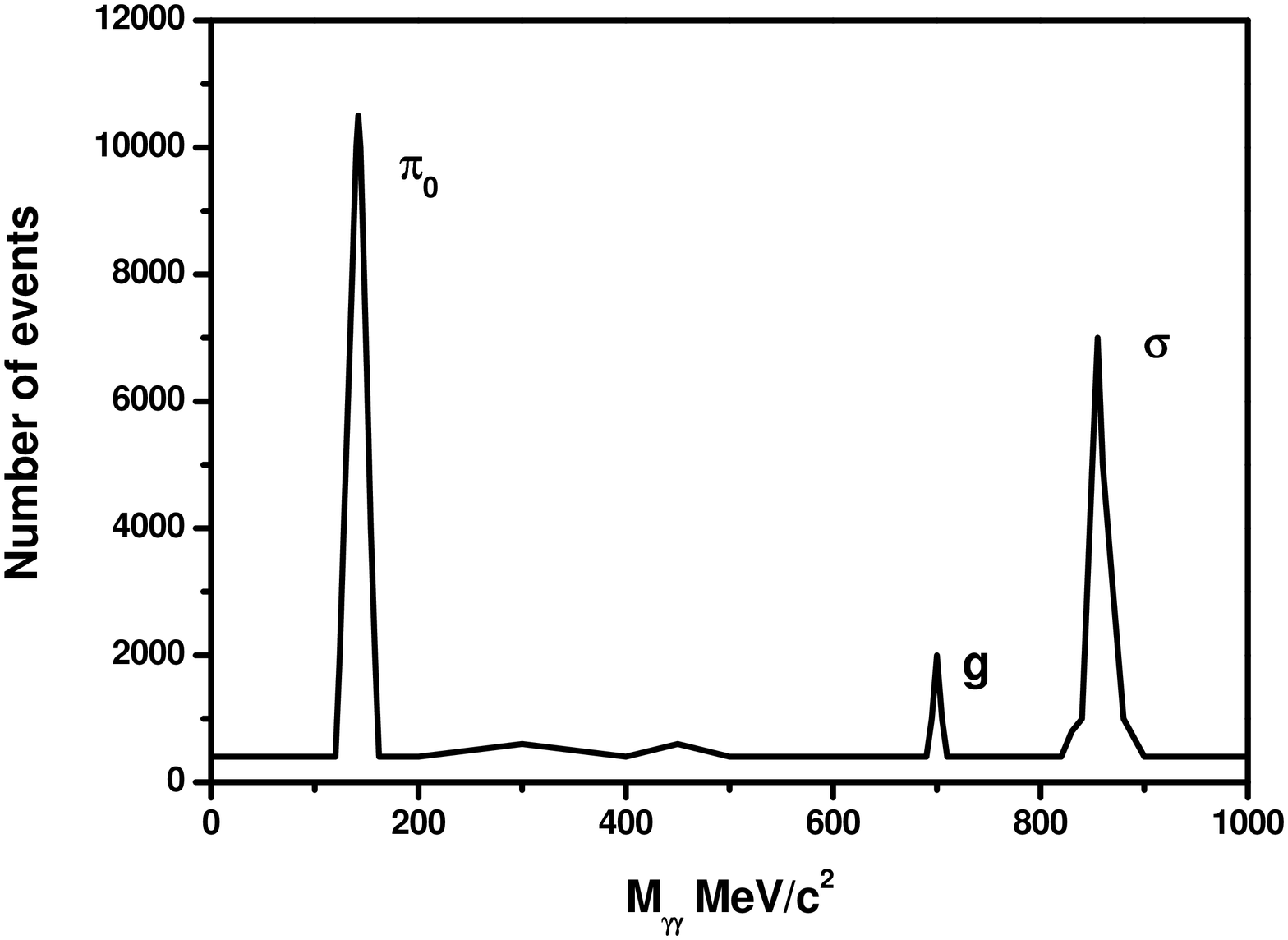,width=5.0cm,angle=270}
\end{minipage} \hskip 1.0cm \begin{minipage}[t]{6.0cm}
\epsfig{file=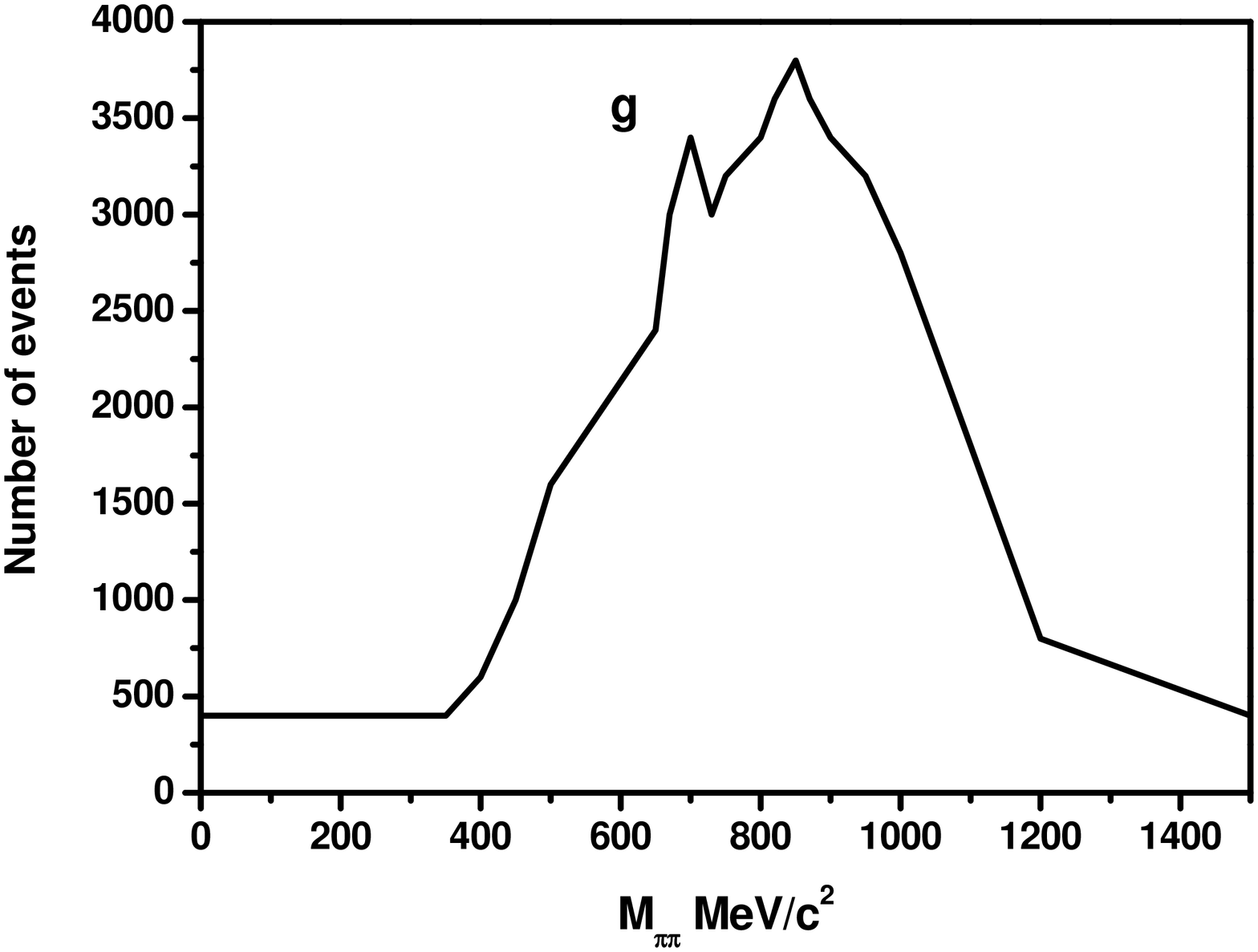,width=5.0cm,angle=270}
\end{minipage}
\caption{\small{Expected fit to the two-photon invariant mass
spectrum and the two-pion invariant mass spectrum in central Pb-Pb
collisions after substraction from the background. The $2\gamma$
decays should allow for a clear separation of $g$ and $\sigma$.
The $2 \pi$ decays will show the $g$ arising above the $\sigma$
background. \label{fits}}} \vskip 0.5cm
\end{figure}

Let me now look into $2\pi$ decays. $\sigma$ and  $g$ are stable
against these type of decays until $m_\sigma$ becomes greater than
$2m_\pi$ for the former, and mixing takes place for the latter.
Thus they cross the phase transition temperature, $T_C$, as stable
states (see Fig.2). This mechanism provides us with a ``time
delay" associated with the cooling after the phase transition.
Thereafter the typical mechanisms for their hadronic decays take
place. Thus, if we were able to measure a  $2\pi$ invariant mass
plot, tagging for this time delay, an enhanced signal would be
observed. If one does not tag, the delay will appear as a
narrowing of the width. One further effect which enhances the
signal is the increase in the number of events due to the larger
population of glueballs in the SCP. Thus the count rate for the
particles will increase and the width of its peaks will narrow. I
expect, therefore, that $g$ will come out from the $\sigma$
background (see Fig. 4).

Let me conclude by stating that I have analyzed the behavior of a
peculiar hadronic state, the scalar glueball, $g$, in a hot
hadronic medium. This state, according to a recent description
\cite{Vento} appears in nature mixed with a scalar meson,
$\sigma$.  I have discussed in physical terms how these particles,
which are created copiously in the strong color Coulomb phase,
behave as the fireball cools down. We have seen that the weak
coupling of $g$ with other hadronic states and the chiral
properties of the $\sigma$  provide them with a well defined
behavior in the plasma as the temperature drops. This behavior is
transferred to its detectable decay products leading to a peculiar
emission of photons and pions  which hints a possible signature
for SCP formation. Finally the large production of the $g$ in the
medium allows for enhanced counting rates compared with the zero
temperature scenario.

\subsection*{Acknowledgments}

I would like to thank K. Langfeld and B.-J. Schaefer for
clarifying information regarding their work. I acknowledge the
hospitality extended to me by the GSI theory group, were this
research was initiated, and in particular the conversations with
Bengt Friman were illuminating. I thank Yu. Kharlov for providing
me with his contribution prior to publication and additional
references. I thank C. Cioffi and D. Treleani for inviting me to
the Trieste conference where this piece of work took shape.

\end{document}